\documentclass[aip,graphicx,reprint,amsmath,amssymb]{revtex4-1}
\draft
\usepackage{dcolumn}
\usepackage{amsmath}
\usepackage{graphicx}
\usepackage{setspace}
\usepackage{amssymb}
\usepackage{dcolumn}
\usepackage{mathrsfs}
\usepackage{bm}

\begin{document}

\preprint{AIP/123-QED}

\title{An integrated single- and two-photon non-diffracting light-sheet microscope} 

\author{Sze Cheung Lau}
\affiliation{Department of Physics, The Hong Kong University of Science and Technology, Clear Water Bay, Kowloon, Hong Kong, China}

\author{Hoi Chun Chiu}
\affiliation{Department of Physics, The Hong Kong University of Science and Technology, Clear Water Bay, Kowloon, Hong Kong, China}

\author{Luwei Zhao}
\affiliation{Light Innovation Technology Ltd., Hong Kong, China}

\author{Teng Zhao} \email{tengzhao@lit.com.hk}
\affiliation{Light Innovation Technology Ltd., Hong Kong, China}

\author{M. M. T. Loy}
\affiliation{Department of Physics, The Hong Kong University of Science and Technology, Clear Water Bay, Kowloon, Hong Kong, China}

\author{Shengwang Du}\email{dusw@ust.hk}
\affiliation{Department of Physics, The Hong Kong University of Science and Technology, Clear Water Bay, Kowloon, Hong Kong, China}
\affiliation{Department of Chemical and Biological Engineering, The Hong Kong University of Science and Technology, Clear Water Bay, Kowloon, Hong Kong, China}

\date{\today}

\begin{abstract}
We describe the apparatus of a fluorescence optical microscope with both single-photon and two-photon non-diffracting light sheets excitation for large volume imaging. With special design to accommodate two different wavelength ranges (visible: 400-700 nm, and near infrared: 800-1200 nm), we combine the line-Bessel  sheet (LBS, for single-photon excitation) and the scanning Bessel beam (SBB, for two-photon excitation) light sheet together in a single microscope setup. For a transparent thin sample where the scattering can be ignored, the LBS single-photon excitation is the optimal imaging solution. When the light scattering becomes significant for a deep-cell or deep-tissue imaging, we use SBB light-sheet two-photon excitation with a longer wavelength. We achieved nearly identical lateral/axial resolution of about 350/270 nm for both imagings. This integrated light-sheet microscope may have a wide application for live-cell and live-tissue three-dimensional high-speed imaging.
\end{abstract}
\maketitle

\section{Introduction}

Light-sheet fluorescence microscopy has been demonstrated for its strength in imaging three-dimensional (3D) live samples as it offers both fast acquisition and significantly reduced phototoxicity as compared the scanning confocal microscopy \cite{GregerRSI2007, StelzerScience2008, Santi2011}. For achieving a high 3D imaging resolution, it is desirable to have the illumination light sheet that is as thin as possible and whose area is as large as possible. However these two requirements contradict each other for the conventional gaussian-beam based light-sheet microscopes (LSM): As the thickness of the light sheet approaches to the diffraction limit of $\lambda/2$, the propagation length (or Rayleigh range) reduces to about one wavelength ($\lambda$) that leads to a very limit field of view. That is partly the reason why the use of gaussian-beam LSM had been focused on large samples with axial resolution of about or above 1 $\mu$m \cite{GregerRSI2007, StelzerScience2008}. A scanning Bessel beam (SBB), with its non-diffraction property, could be implemented to extend the the sheet area and obtain both lateral and axial resolution approaching the diffraction limit, but its strong side lobes induces additional off-plane phototoxicity \cite{BesselBeamLSM01}. Therefore, a SBB LSM is often used as excitation for nonlinear microscopies, such as two photon microscopy \cite{BesselBeamTwoPhoton01, BesselBeamTwoPhoton02} and stimulated emission depletion microscopy \cite{BesselBeamSTED}. For single-photon LSM, E. Betzig \textit{et al.} invented the lattice LSM (LLSM) in which non-diffracting light sheets are realized with two-dimensional  lattice pattern structures \cite{LatticeLSM}.  Recently, we demonstrated a simple and robust method to create non-diffracting ultrathin line Bessel sheets (LBS) for single-photon LSM \cite{LBS}.

It is of practical importance to have a single microscope system that can be operated in both single- and two-photon excitation modes. For the single-photon excitation, LLSM or LBS is the preferable choice but its direct 2D sheet configuration, which spreads the light energy on a large area and results in weak intensity, is not suitable for two-photon nonlinear process that requires a high instantaneous intensity. The SBB is the optimal solution for the two-photon LSM. Building the LLSM/LBS and SBB together in one system remains a technical challenge because of their incompatible modes. In this Article, we demonstrated a novel setup that combines this single-photon LBS and two-photon SBB light sheet into a single microscope system. We achieved a near isotropic resolution of around 300nm for both single and two-photon imaging modes. In the single-photon LBS, we acquires 3D images of cells or superficial layer of tissues with ease, while the two-photon SBB can be used for imaging deep tissue or small animals like zebrafish for better signal to noise ratio (SNR).

\section{Optical Setup}
\begin{figure*}
\includegraphics[width=17cm]{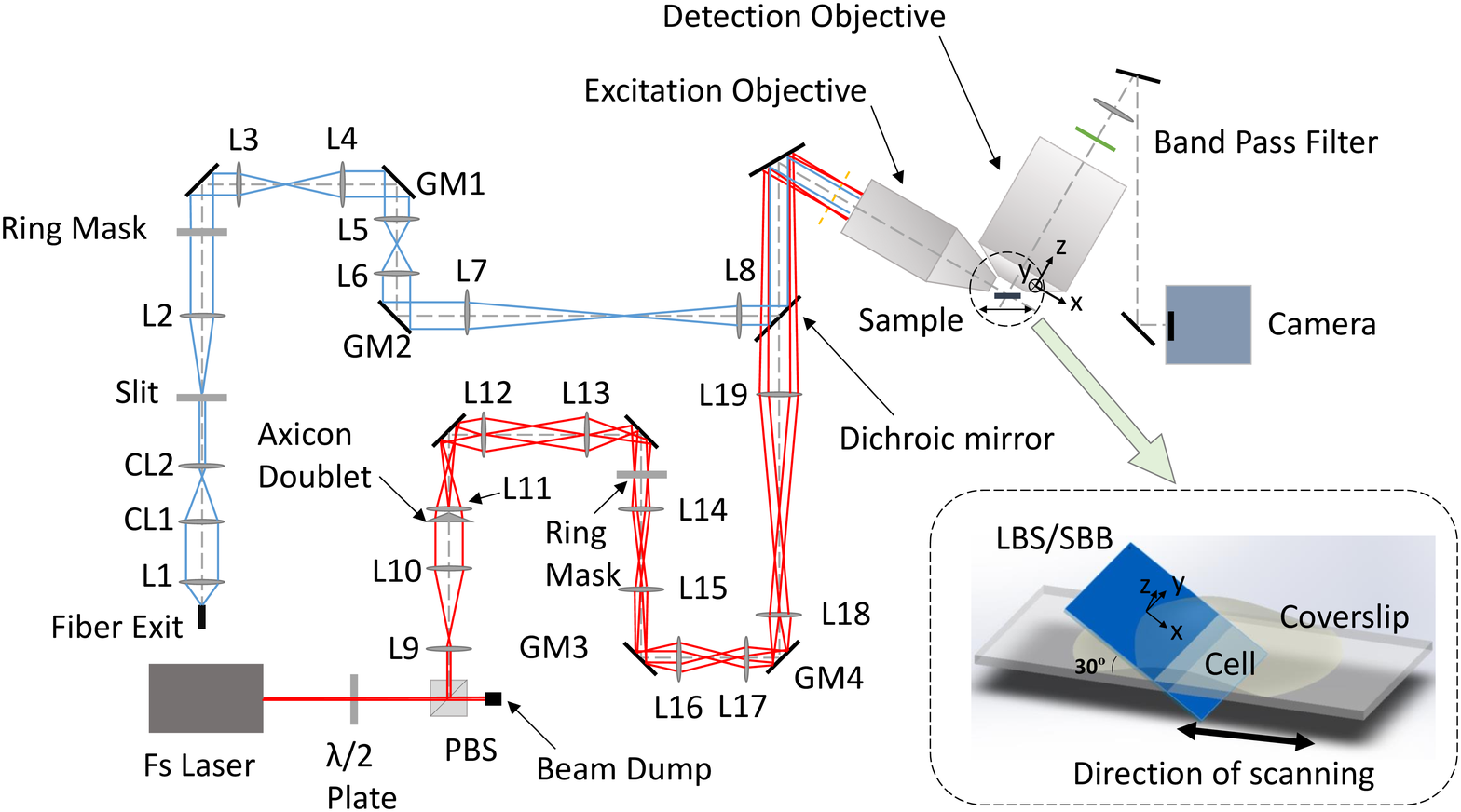}
\caption{\label{fig:Setup} Optical setup of the integrated single-photon LBS and two-photon SBB non-diffracting light-sheet microscope. Blue: LBS single-photon light sheet (488 nm). Red: SBB two-photon light sheet (800 nm). All lenes used in the system are achromatic doublets to minimize the chromatical abberations due to the large wavelength span. L1 to L8 and CL1 and CL2 are all designed for visitable light 400-700 nm while lenses L9 to L19 are designed for 650-1050 nm. In the coordinate system, x axis is along the sheet beam propagation direction, and z direction is along the imaging optical axis.}
\end{figure*}

The experimental schematic of the integrated non-diffracting light-sheet microscope is shown in Fig.~\ref{fig:Setup}, where the single-photon LBS (488 nm, blue traces) and two-photon SBB (680-1080 nm tuneable, red traces) light sheets are combined at a dichroic mirror (longpass cutoff at 680nm). The optical system design takes into account the large wavelength difference and the mode compatibility. Lenses L1 to L8 are all achromatic doublets designed at visible wavelengths (400-700 nm) while L9 to L19 are achromatic doublets for near infrared beams (650-1050 nm). The lateral positions of LBS and SBB can be adjusted individually by galvo mirrors GM2 and GM4, while the axial position of the SBB can be fine tuned by carefully shifting L19 along beam axis. With all these freedoms of adjustments, single-photon LBS and two-photon SBB can be both perfectly aligned to the focal plane of the detection objective. As shown in Fig.~\ref{fig:Setup} and its inset, the xyz coordinate system is chosen as the following: the sheets are on x-y plane with x axis along the beam propagation direction, and  z direction is along the imaging optical axis.

\subsection{LBS Single-Photon Light Sheet}

The blue rays in Fig.~\ref{fig:Setup} illustrate the optical layout of LBS single-photon light sheet. We use a diode laser (488 nm, Coherent OBIS) output as the single-photon excitation light source. After passing through an acousto-optical tunable filter (Gooch \& Housego, 400-700 nm) for intensity modulation, the laser light is coupled to an optical single-mode fiber (not shown). The collimated light from the fiber exit is shaped by a pair of orthogonal cylindrical lenses (CL1 and CL2) to compress in one direction so that the most of the light passes through a single slit with a width of 0.2 mm. The Fourier transform of the slit is then cropped by a ring mask (with inner and outer radiuses of 1.973 mm and 0.657 mm, respectively) for achieving the non-diffracting property. At this point, an image of two collinear and truncated lines pass through the mask. This image is conjugated with two galvo-mirrors scanning in x (GM1) and z (GM2) directions, and finally to the back focal plane of excitation objective (Light Innovation Technology Ltd, LS29 ApoPlan LWD 29x 0.55NA, water immersion) by pairs of relay lenses (L7-L8). The excitation objective transforms the pattern at its back focal plane to produce the final light sheet with center thickness less than 400 nm. GM1 scans the light sheet in the focal plane of the detection objective such that a much larger field of view can be covered, and GM2 aligns the light sheet precisely to the focal plane of the detection objective. The details of LBS is also described in Ref. \cite{LBS}.

\subsection{SBB Two-Photon Light Sheet}

The red rays in Fig.~\ref{fig:Setup} illustrate the optical layout of SBB two-photon light sheet. Here we work with a femtosecond laser (800 nm, 2.9 W, Coherent Chameleon Ultra I).  A broad-band half-wave plate (680-1080 nm) in the beam path and is rotated to adjust the amount of laser power reflected by the polarizing beam splitter (PBS) while the rest of the energy is dumped to a heat-dissipating beam dump. The reflected laser beam is then expanded by a telescope lens pair and shaped by an axicon (Thorlabs, AX251-B) closely placed in front of the lens L11 to produce a ring image. This images is cropped by a ring mask (with inner and outer radiuses of 7.17 mm and 6.023 mm, respectively) to filter out the unwanted components produced by the imperfection of the axicon. This ring image is conjugated to another set of galvo-mirrors (GM3 and GM4) and further relayed by the lens pair L18 and L19, then finally projected to the back focal plane of excitation objective, and produce the SBB at the focus of the excitation objective which is carefully aligned to the focal plane of the detection objective, and overlaps with LBS. The SBB is scanned by GM3 to form the light sheet for imaging.

\subsection{Fluorescence Imaging System}

In the fluorescence imaging part, we use a high-NA detection objective (Nikon, CFI APO LWD, NA=1.1, water immersion) which is placed 90$^o$ to the excitation objective and 60$^o$ to the sample plane. We define the z axis of the system to be along the optical axis of the detection objective, and the x axis is along that of the excitation objective. A set of band pass filters are placed after the detection objective to select the fluorescence signal. The final image is formed by a tube lens and detected by a sCMOS camera (Hamamatsu Flash 4.0 V3).

\section{Characterization}

\begin{figure}
\includegraphics[width=8.5cm]{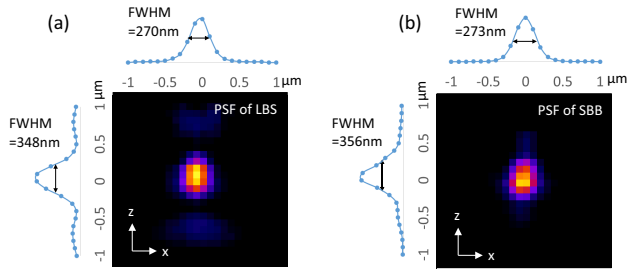}
\caption{\label{fig:PSF} Measurements of the PSFs. The y projections on z-x plane of the 30-nm bead images for (a) LBS single-photon and (b) SBB two-photon microscopies.}
\end{figure}

To characterize the system performance, we use 30nm-diameter fluorescence beads for measuring the 3D point spread functions (PSFs) of both LBS and SBB microscopies. The beads were coated on cover slip and excited by 488nm and 800nm laser light for LBS and SBB, respectively. The y projections on z-x plane of the bead images are displayed in Fig.~\ref{fig:PSF}. The y-direction, which is identical to the x-direction, is not shown here. The full widths at half maximum (FWHM) of the PSFs are 348 nm (axial) and 270nm (lateral) in LBS single-photon microscopy, and 356 nm (axial) and 273 nm (lateral) in SBB two-photon microscopy. Thus, we achieve nearly the same 3D resolution for both LBS single-photon and SBB two-photon images.

\begin{figure}
\includegraphics[width=8.5cm]{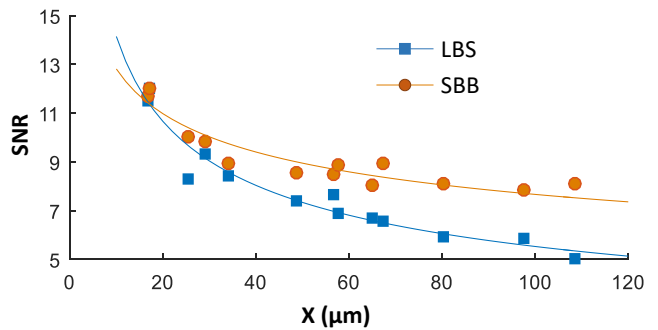}
\caption{\label{fig:SNR}  SNR as a function of depth along x-axis for LBS single-photon and SBB two-photon LSMs. The solid curves are exponential decay with best fitted decay coefficients 0.016/$\mu$m and 0.009/$\mu$m for LBS and SBB, respectively.}
\end{figure}

The advantage of the two-photon microscopy is that its longer wavelength excitation allows deeper sample imaging with reduced scattering, as compared to the single-photon microscopy. We compare the imaging depth of the LBS single-photon and SBB two-photon LSMs using the same sample of yeasts with a part of DNA strands labeled with green fluorescent protein (GFP) immersed in 6\% agarose which is a good analogy of the structures of a thick biological tissue. The SNR at different penetration depth of both light sheets are plotted in Fig.~\ref{fig:SNR}. The decay coefficients for LBS and SBB are 0.016/$\mu$m and 0.009/$\mu$m respectively. This result indicates that the SBB two-photon light sheet is able to image (at the same SNR) much deeper into the yeast-agarose sample than the LBS single-photon light sheet.

\section{Examples}

\begin{figure}
\includegraphics[width=8.5cm]{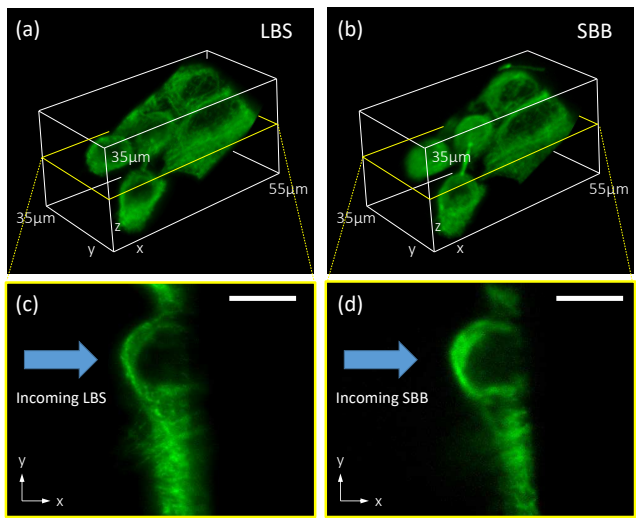}
\caption{\label{fig:ThinSample} Imaging a thin sample of microtubules of HT22 cells.  (a) and (b) are the 3D images of cells obtained using LBS single-photon and SBB two-photon LSMs, respectively. (c) and (d) are the x-y cross section views along the plane as marked by yellow lines in (a) and (b), respectively. Scale bars: 10 $\mu$m.}
\end{figure}

\begin{figure}
\includegraphics[width=8.5cm]{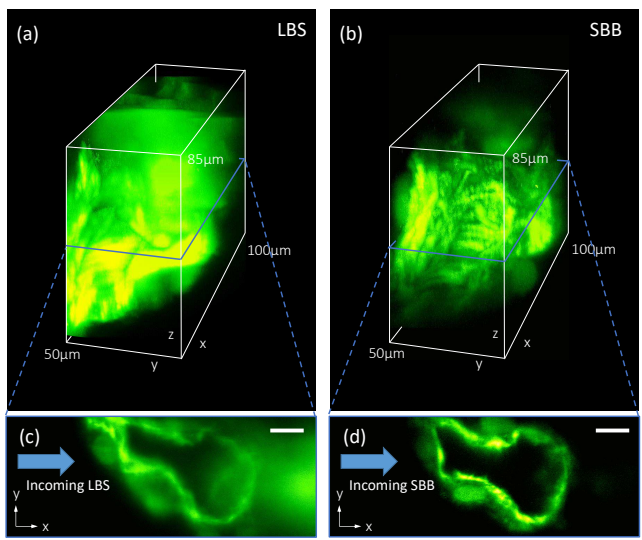}
\caption{\label{fig:ThickSample} Images of a GFP tagged zebrafish heart (myocardial structures) in a 48hpf Zebrafish embryo using (a) LBS single-photon and (b) SBB two-photon LSM. (c) and (d) are the x-y cross section views along the plane indicated by the blue lines in (a) and (b), respectively. Scale bars :10 $\mu$m.}
\end{figure}

As examples of application, we first image Alexa488 tagged microtubules in fixed HT22 cells cultured on a coverslip to show the capability of both LBS and SBB light sheets in thin samples, as shown in Fig.~\ref{fig:ThinSample}. The cells form a monolayer on the surface of a coverslip with 10 $\mu$m depth along the x-axis. For this thin samples, both single-photon and two-photon light sheets imaging provide nearly the same resolution and SNR. Here the laser light power for LBS is 20 $\mu$W, while for SBB is 300 mW. Obviously for imaging thin samples, LBS single-photon LSM is the optimal solution with much weaker excitation laser power.

To explore the performance of LBS and SBB in deep tissue imaging, we image the GFP tagged zebrafish heart (myocardial structures) in a 48hpf Zebrafish embryo. Fig.~\ref{fig:ThickSample} shows the 3D images and the x-y plane cross-sections of the myocardial structures up to 100 $\mu$m deep beneath the surface of the embryo, taken by the LBS single-photon and SBB two-photon LSM, respectively. Obviously, in this scattering thick samples, LBS single-photon LSM suffers significantly greater background and thus much reduced SNR in comparison to the SBB two-photon LSM, where in the later technique, its longer excitation wavelength reduces the scattering and thus maintains the shape of light sheet for a longer propagation distance than that in the LBS. Meanwhile, the scattered laser light becomes too weak to excite fluorophores that are outside the light sheet. As a result, The out-of-sheet background noise is also significantly reduced in the two-photon LSM.

\section{Summary and Discussion}

In summary, we have developed a LSM integrated with both LBS single-photon and SBB two-photon excitations. For imaging thin samples with negelable  scattering, both LBS and SBB give nearly identical 3D axial/lateral resolution of about 350 nm/270 nm and high SNR, but LBS is the preferred solution because of its much weaker laser power consumption and easier maintenance, as confirmed by imaging microtubules of HT22 cells. For thick samples with significant single-photon scattering, the SBB two-photon LSM provides much better imaging quality, as shown in the Zebrafish Embryo heart images.

Although in this paper we demonstrate the capability of the system using single wavelengths LBS at 488 nm, multiple laser lines have been intergraded into the system for multi-color single photon florescent imaging. It is worthwhile to mention that, because of the wider application of single-photon fluorescence imaging, various good single-photon fluorophores are readily available with known spectra and other properties, but this is not the case in two-photon excitation. Efficient single-photon fluorophores are not necessarily good in two-photon. To image samples in two-photon with good contrast, more research is needed to be done to choose the best fluorophore and most suitable wavelength. To obtain the same contrast, two-photon laser power needs to be tens of thousands or hundreds of thousands stronger than single-photon laser power. In this high-power setting, photobleaching becomes serious, and even the heating effect could cause damage or degradation on the sample. In this sense, our developed microscope not only extends the capacity of a LSM, but also provides a testing platform for two-photon fluorophores which can be compared directly with its single-photon excitation of the same imaging area in the same setup.

\begin{acknowledgements}
The work was supported by a grant from the Offices of the Provost, VPRG and Dean of Science, HKUST (Project No. VPRGO12SC02), and the Hong Kong Research Grants Council (Project No. C6030-14E). We acknowledge Aifang Cheng and Claire Shuk-Kwan Lee at Division of Life Science, HKUST for providing the biological specimens. 
\end{acknowledgements}

\end{document}